# Experimental discovery of Weyl semimetal TaAs

B. Q. Lv[1,§], H. M. Weng[1,2,§], B. B. Fu[1], X. P. Wang[2,3,1], H. Miao[1], J. Ma[1], P. Richard[1,2], X. C. Huang[1], L. X. Zhao[1], G. F. Chen[1,2], Z. Fang[1,2], X. Dai[1,2], T. Qian[1,*], and H. Ding[1,2,*]

[1] *Beijing National Laboratory for Condensed Matter Physics and Institute of Physics, Chinese Academy of Sciences, Beijing 100190, China*
[2] *Collaborative Innovation Center of Quantum Matter, Beijing, China*
[3] *Department of Physics, Tsinghua University, 100084, Beijing, China*

Abstract

Weyl semimetals are a class of materials that can be regarded as three-dimensional analogs of graphene breaking time reversal or inversion symmetry. Electrons in a Weyl semimetal behave as Weyl fermions, which have many exotic properties, such as chiral anomaly and magnetic monopoles in the crystal momentum space. The surface state of a Weyl semimetal displays pairs of entangled Fermi arcs at two opposite surfaces. However, the existence of Weyl semimetals has not yet been proved experimentally. Here we report the experimental realization of a Weyl semimetal in TaAs by observing Fermi arcs formed by its surface states using angle-resolved photoemission spectroscopy. Our first-principles calculations, matching remarkably well with the experimental results, further confirm that TaAs is a Weyl semimetal.

§ These authors contributed equally to this work.
* Corresponding authors E-mail: tqian@iphy.ac.cn, dingh@iphy.ac.cn

Although the subjects of high-energy physics and condensed matter physics are very different, they sometimes share the same ideas. The most famous examples are the concepts of spontaneously broken symmetry and the Higgs mechanism. Recently, the basic concepts of Dirac and Weyl fermions (1, 2) in quantum field theory have also been applied to condensed matter systems (3-7), in particular following recent progress in the field of topological insulators (8, 9). From the Dirac equation, massless Dirac fermions are described by the crossing of two spin-degenerate bands. A Dirac semimetal possesses such four-fold degenerate Dirac nodes at the Fermi level ($E_F$) and has been realized recently (10-16). Weyl fermions, which have not yet been discovered in high-energy physics, can be realized as an emergent phenomenon by breaking either time-reversal symmetry or inversion symmetry in Dirac semimetals, where a Dirac node can be regarded as two Weyl nodes with opposite chirality overlapping each other (Fig. 1A). A material with separated Weyl nodes is called a Weyl semimetal (WSM) (3-7).

Due to the no-go theorem (17, 18), the Weyl nodes in a WSM must come in pairs with opposite chirality. Separated Weyl nodes are topologically stable (19-20) because any perturbation respecting the translational symmetry can only shift but not annihilate them (21, 22). This is different from topological insulators whose topological properties are protected by the energy gap in the bulk state. Therefore, a WSM can be viewed as a new type of topological non-trivial phase other than the $Z_2$ topological insulators, making WSMs a good platform for studying and manipulating novel topological quantum states with a promising application potential.

A hallmark of a WSM is the existence of Fermi arcs on the surface (3-5), which is a direct consequence of separated Weyl nodes with opposite chirality, with the two ending points of the Fermi arc coinciding with the Weyl-node projections on the surface (Fig. 1B). Another unique property of a WSM is the chiral anomaly (23-25), which implies the apparent violation of charge conservation and leads to interesting transport properties, such as negative magneto-resistance, chiral magnetic effects and anomalous Hall effect (26-28).

In this work, we report the observation of Fermi arcs on the surface of noncentrosymmetric and nonmagnetic material TaAs using angle-resolved photoemission spectroscopy (ARPES). Unlike the previously proposed WSMs which are usually complex materials tailored by fine tuning methods (29-34), TaAs is predicted to be a WSM in its natural state (35, 36). The removal of the spin degeneracy of the bands is induced by the lack of inversion symmetry in the crystal structure rather than by the breaking of the time reversal symmetry. Such realization of the WSM phase in nonmagnetic materials allows direct observation of the Fermi arcs by ARPES because the complexity caused by a magnetic domain structure is absent.

The crystal structure of TaAs shown in Fig. 1C has the nonsymmorphic space group $I4_1md$. Due to the lack of inversion symmetry, first-principles calculations predict that TaAs is a time-reversal invariant three-dimensional (3D) WSM with a dozen pairs of Weyl nodes in the Brillouin zone (BZ) (35, 36). To prove this experimentally, we investigate the electronic structure of TaAs single crystals using ARPES. The core level spectrum in Fig. 1E shows the characteristic peaks of Ta and As elements, confirming the chemical composition of TaAs samples. It is important to notice that the As $3d$ core levels have two sets of spin-orbit doublets, while the Ta $4f$ core levels have only one set of doublets, suggesting that the cleaved surface is As terminated. The cleaved (001) surfaces in our measurements are very flat at the millimeter scale (Fig. 1, F and G), which is larger than the 30×20 μm$^2$ spot size of the incident light in our ARPES experiments and thus a single surface domain can be measured. High-quality ARPES data of highly dispersive and well-defined quasiparticle peaks are obtained (Fig. 1H), enabling us to precisely determine the band structure and the Fermi surface (FS).

To investigate the electronic structure in the 3D BZ, we carried out photon energy dependent ARPES measurements. To our surprise, almost all the observed band dispersions do not show any noticeable change with varying the incident photon energy over a wide range (20 – 200 eV) (some examples are shown in Fig. 2, A to E), indicating that they are non-dispersive along $k_z$. To understand the two-dimensionality of the experimental band structure, we carried out first-principles band structure calculations for a (001) oriented seven-unit-cell-thick slab (37) with top and bottom surfaces terminated by As and Ta layers, respectively (Fig. 2, G and H). The experimental band dispersions are remarkably well reproduced by the calculated surface state band structure with the As termination, in agreement with the conclusion from our core level data. The absence of bulk bands can be understood from our calculations that the bulk bands have vanishing spectral weight within the topmost unit cell, where most of the surface-sensitive ARPES signal origins.

The absence of bulk bands in the low-photon-energy ARPES measurements enables us to compare directly the measured surface states and their FSs with the calculated results with great precision. The consistency between calculations and experiments is further confirmed by the overall electronic structure along high-symmetry lines (Fig. 2, I to K). As the four-fold symmetry is broken on the cleavage (001) surface, the surface bands are predicted to be anisotropic along $\overline{\Gamma}$-$\overline{X}$ and $\overline{\Gamma}$-$\overline{Y}$. Indeed, such anisotropic features are clearly observed in our ARPES data (Fig. 2J). The excellent consistency between experiments and calculations supports our FS assignments and our conclusion on WSM.

The topological nature of the WSM requires each Fermi arc to start and end at the surface projections of two Weyl nodes with opposite chirality. For systems with multiple pairs of Weyl nodes, the surface connection pattern is not uniquely determined by the bulk band structure and can be modified by changing the surface condition and the chemical potential. On the TaAs (001) surface, there are topologically nontrivial surface states forming Fermi arcs as well as normal surface states forming FS pockets. The appearance of normal surface states can generate a complicated FS structure on the surface. It is remarkable that the allover surface FS pattern, such as the two "cross"-like FSs centered at both $\overline{X}$ and $\overline{Y}$, and the "horseshoe"-like FSs around the loci of the projected Weyl nodes W1, matches so well in both ARPES measurements and theoretical calculations (Fig. 3, A and B).

Before discussing the detailed connection pattern of the FS, we would like to give a general and simple argument on the existence of Fermi arcs in the surface BZ. As demonstrated in Fig. 3C, it is obvious that a closed FS can only cross an arbitrary $k$-loop in surface BZ an even number of times. Only an open Fermi arc can possibly cross this loop an odd number of times. Therefore, if we can find a specific loop containing a total odd number of FS crossings, Fermi arcs must exist. This is a sufficient condition for the existence of Fermi arcs. For TaAs, we choose the $\overline{\Gamma}$-$\overline{X}$-$\overline{M}$-$\overline{\Gamma}$ and $\overline{\Gamma}$-$\overline{Y}$-$\overline{M}$-$\overline{\Gamma}$ closed *k* paths shown in Fig. 3D as reference loops. Through careful inspection of ARPES dispersion as discussed in the Appendix B, we determined that the surface FSs cross the closed loop $\overline{\Gamma}$-$\overline{X}$-$\overline{M}$-$\overline{\Gamma}$ seven times, and five times for the $\overline{\Gamma}$-$\overline{Y}$-$\overline{M}$-$\overline{\Gamma}$ loop, thus providing a direct experimental evidence of the existence of Fermi arcs on the (001) surface of TaAs.

Next we determine the Fermi arcs and their topology with the help of first-principles calculations. The precise agreement with the experimental observation of the surface band structure gives us confidence in going beyond the resolution of the present ARPES experiment. The calculated surface FS using extra fine *k*-point sampling around the Weyl-node projections W2 and W1 along the $\overline{\Gamma}$-$\overline{Y}$ direction are shown in Fig. 4, A and B. The situation along $\overline{\Gamma}$-$\overline{X}$ is nearly the same and the following discussion can be applied to both. Through detailed examination of ARPES measurements and theoretical calculations shown in Fig. 4 as discussed in the Appendix B, we summarize the Fermi arcs observed on the (001) surface of TaAs in Fig. 4C by ignoring the trivial FS pockets. It is clear that the *a*1 arc connects the W2 and W1 nodes and the *a*5 arc connects two W2 nodes. Such connections are consistent with their topological charges, *i.e.* ±2 for W1 and ±1 for W2. Furthermore, the mirror Chern number of the mirror plane passing through $\overline{\Gamma}$-$\overline{Y}$ is 1, which imposes the number of edge states crossing $E_F$ to be odd (35). The connection pattern obeys this rule since only the *a*5 arc crosses $\overline{\Gamma}$-$\overline{Y}$ one time.

In summary, we have observed Fermi arcs on the (001) surface of TaAs by using ARPES. The surface FS is found to cross a closed *k*-loop on the (001) surface BZ an odd number of times, which gives a strong evidence for an odd number of Weyl nodes enclosed inside the loop. Our first-principles calculations reproduce almost every detail of the experimental measurements, including the band dispersion and the surface FS. The shape of the Fermi arcs and their connectivity are further identified and clearly shown for the first time. Our study confirms that we have discovered a Weyl semimetal in TaAs.

Note Added: After the completion of this work, we subsequently performed experiments investigating the bulk electronic structure of TaAs and observed the existence of Weyl nodes, another characteristic of Weyl semimetals. The latter observation, which corroborates the finding in this paper, was reported in Ref. [38].

At the time this work was made public through posting on arXiv as eprint arXiv:1502.04684, a similar work, carried out independently by another group, also became public as eprint arXiv:1502.03807. A revised version of that paper was recently published: see Ref. [39].

**Acknowledgments:** This work was supported by the Ministry of Science and Technology of China (No. 2013CB921700, No. 2015CB921300, No. 2011CBA00108, and No. 2011CBA001000), the National Natural Science Foundation of China (No. 11474340, No. 11422428, No. 11274362, and No. 11234014), and the Chinese Academy of Sciences (No. XDB07000000).

## APPENDIX A: Materials and Methods

A1:   Sample growth and preparations

High quality TaAs single crystals were grown by the chemical vapor transport method. A polycrystalline sample was filled in a quartz ampoule using iodine as transporting agent of 2mg/cm$^3$. After evacuating and sealing, the ampoule was kept at the growth temperature for three weeks. Large polyhedral crystals with dimensions up to 1.5 mm were obtained in a temperature field of $\Delta T$ = 1150 °C-1000 °C. The as-grown crystals were characterized by X-ray diffraction using PANalytical diffractometer with Cu K$\alpha$ radiation at room temperature. The crystal growth orientation is determined by single-crystal X-ray diffraction and the average stoichiometry was determined by energy-dispersive X-ray spectroscopy. TaAs consists of alternating stacking of Ta and As layers and has a NbAs type body-centered-tetragonal structure (40). The corresponding space group is $I4_1md$ and the lattice parameters are $a = b$ = 3.4348 Å and $c$ = 11.641 Å. The adjacent TaAs layers are rotated by 90° and shifted by $a$/2.

A2:   Angle-resolved photoemission spectroscopy (ARPES) measurements

ARPES measurements were performed at the "Dreamline" beamline of Shanghai Synchrotron Radiation Facility (SSRF) with a Scienta D80 analyzer. The samples were cleaved *in situ* and measured at 25 K in a vacuum better than $5 \times 10^{-11}$ Torr. The energy resolution was set at 15 meV for the Fermi surface mapping and the angular resolution was set at 0.2°. The ARPES data were collected using linearly horizontal-polarized lights with a vertical analyzer slit.

A3:   First-principles calculations of the band structure

First-principles calculations were performed using the OpenMX (41) software package. The pseudo atomic orbital basis set is chosen as Ta9.0-s2p2d2f1 and As9.0-s2p2d1. The exchange-correlation functional within generalized gradient approximation parameterized by Perdew, Burke, and Ernzerhof has been used (42). The optimized crystal structure is used. To calculate the (001) surface state of TaAs, we have built a superlattice composed by a seven-unit-cell-thick (001) oriented slab and a vacuum layer of 12 Å. The slab has its top and bottom surfaces terminated by As and Ta layer, respectively. To identify the surface state, the weight on the outmost one unit-cell is calculated for each wave function.

## APPENDIX B: Detailed Analysis of Band Structures

B1: Inspection of Fermi crossings along closed loops

The two closed loops shown in Fig. 3D are $\bar{\Gamma}$ - $\bar{X}$ - $\bar{M}$ - $\bar{\Gamma}$ and $\bar{\Gamma}$ - $\bar{Y}$ - $\bar{M}$ - $\bar{\Gamma}$. Each one encloses three Weyl nodes predicted theoretically. Note that the Weyl node W1 is doubly degenerate as projected to the (001) surface Brillouin zone. Here we elaborate how we identify Fermi crossings for the two closed loops. It is clear that there is no Fermi crossing along $\bar{\Gamma}$ - $\bar{M}$ (Fig. 2, I and J). We identify three Fermi crossings along $\bar{\Gamma}$ - $\bar{X}$ at $k_x$ = 0.35, 0.44 and 0.945 π/a, respectively (Fig. 2, A to E). The first two crossings at $k_x$

= 0.35 and 0.44 $\pi/a$ are clear. There are three bands identified near $E_F$ around $\overline{X}$. While the outermost band crosses $E_F$ at $k_x$ = 0.945 $\pi/a$, the two inner bands curl down below $E_F$, leading to the feature $k_x$ = 0.975 $\pi/a$. The calculated bands in Fig. 2G also show that these two inner bands are hole-like but do not cross $E_F$. We identify four Fermi crossings along $\overline{X}$-$\overline{M}$ (Fig. 3, E and F). The first two crossings close to $\overline{X}$ come from one electron-like band with a bottom just below $E_F$, in agreement with the calculations. The third and fourth ones at $k_y$ ~ 0.35 $\pi/a$ are from two nearly degenerate bands, which are separated along cut $C2$ at the higher binding energy (Fig. 3, G and H). Along $\overline{\Gamma}$-$\overline{Y}$, the bands are similar to those along $\overline{\Gamma}$-$\overline{X}$ and there are three Fermi crossings. Along $\overline{Y}$-$\overline{M}$, there are two Fermi crossings (Fig. 3, I and J).

B2: Determination of Fermi arcs topology and connection pattern

As seen from Fig. 4, A to E, there is one Fermi arc noted as $a1$ connecting W1 and W2 along $\overline{\Gamma}$-$\overline{Y}$. Experimentally, $a1$ and $a2$ are nearly degenerate (Fig. 4, D and E). To distinguish them, we measured the band dispersion along cut $C4$ (Fig. 4, G and H), where $a1$ and $a2$ are well separated around a binding energy of 0.4 eV. Although the calculated $a1$ line has relatively smaller weight at the surface around $E_F$, more surface weight is recovered at higher binding energy as shown in Fig. 4H. The horseshoe-like arc $a5$, connecting the adjacent W1 points with opposite chirality on either side of $\overline{\Gamma}$-$\overline{Y}$, is clearly identified in both experiments and calculations. In addition, one can see three other arc-like lines around W1, namely $a2$, $a3$, and $a4$. The $a2$ and $a4$ lines are also observed in our ARPES experiments but they are connected to each other, thus revealing their trivial nature. The $a3$ line, or equivalently $b3$, which is not seen in experiment (Fig. 4I), is mainly a bulk state. Our calculation along $C5$ (Fig. 4J), also shows that $b3$ has small weight at the surface at all binding energy, further supporting its bulk nature. Its spectral weight on the surface is increased slightly only when approaching the surface band $b4$ near $E_F$. This is the reason why $b3$ appears in the calculated Fermi surface in Fig. 4F. W2 has a topological charge of 1 and is connected to W1 by the $a1$ arc. The arc-connecting pattern around W2 is indicated in Fig. 4K. The surface state contributing to the $a1$ arc crosses $E_F$ along C6 and curls down below $E_F$ along C7 (Fig. 4, L and M).

B3: Difference of Fermi arcs in Weyl semimetal and Dirac semimetal

Here we draw attention to the difference between a Fermi arc on the surface of a Weyl semimetal and a Fermi arc on the surface of a Dirac semimetal, as in the case of $Na_3Bi$ (10, 12). Since a Dirac node is composed by two "kissing" Weyl nodes with opposite chirality (3, 4), it serves as both Fermi arc "source" and "sink" at the same time. There are always two Fermi arcs connected to one projected Dirac node. Theoretically, these two arcs just touch each other instead of forming a continuing closed Fermi surface pocket (10). However, this raises a serious concern on how to unambiguously identify two touching arcs or one closed Fermi surface pockets in ARPES experiments.

B4: Evolution of band dispersions while sliding through W1

Figure 5 show the evolution of band dispersion while sliding through the Weyl nodes W1. The outer band $a5$ approaches the inner one ($a4$) from $D1$ to $D3$, which is accompanied by the suppression of spectral weight, and become indistinguishable along $D4$. The band $a4$ along $D1$ smoothly evolves into $a2$ along $D5$ while sliding through W1, indicating the $a2$ and $a4$ are connected to each other, thus revealing their trivial nature.

**References**

1. H. Weyl, *Electron and Gravitation. I*, Z. Phys. **56**, 330 (1929).

2. G. E. Volovik, *The Universe in a Helium Droplet*, Oxford University Press (2009).

3. X. Wan, A. M. Turner, A. Vishwanath, and S. Y. Savrasov, *Topological Semimetal and Fermi-Arc Surface States in the Electronic Structure of Pyrochlore Iridates*, Phys. Rev. B **83**, 205101 (2011).

4. L. Balents, *Weyl Electrons Kiss*, Physics **4**, 36 (2011).

5. G. Xu, H. Weng, Z. Wang, X. Dai, and Z. Fang, *Chern Semimetal and the Quantized Anomalous Hall Effect*, Phys. Rev. Lett. **107**, 186806 (2011).

6. A. M. Turner and A. Vishwanath, *Beyond Band Insulators: Topology of Semi-Metals and Interacting Phases*, arXiv:1301.0330.

7. O. Vafek and A. Vishwanath, *Dirac Fermions in Solids: From High-$T_c$ Cuprates and Graphene to Topological Insulators and Weyl Semimetals*, Annu. Rev. Condens. Matter Phys. **5**, 83 (2014).

8. M. Z. Hasan and C. L. Kane, *Colloquium: Topological Insulators*, Rev. Mod. Phys. **82**, 3045 (2010).

9. X.-L. Qi and S.-C. Zhang, *Topological Insulators and Superconductors*, Rev. Mod. Phys. **83**, 1057 (2011).

10. Z. Wang, Y. Sun, X.-Q. Chen, C. Franchini, G. Xu, H. Weng, X. Dai, and Z. Fang, *Dirac Semimetal and Topological Phase Transitions in $A_3Bi$ (A = Na, K, Rb)*, Phys. Rev. B **85**,195320 (2012).

11. Z. K. Liu, B. Zhou, Y. Zhang, Z. J. Wang, H. M. Weng, D. Prabhakaran, S-K. Mo, Z. X. Shen, Z. Fang, X. Dai, Z. Hussain, and Y. L. Chen, *Discovery of a Three-Dimensional Topological Dirac Semimetal, $Na_3Bi$*, Science **343**, 864 (2014).

12. S. Y. Xu, C. Liu, S. K. Kushwaha, R. Sankar, J. W. Krizan, I. Belopolski, M. Neupane, G. Bian, N. Alidoust, T.-R. Chang, H.-T. Jeng, C.-Y. Huang, W.-F. Tsai, H. Lin, P. P. Shibayev, F.-C. Chou, R. J. Cava, and M. Z. Hasan, *Observation of Fermi Arc Surface States in a Topological Metal*, Science **347**, 294 (2015).

13. Z. Wang, H. Weng, Q. Wu, X. Dai, and Z. Fang, *Three-dimensional Dirac Semimetal and Quantum Transport in $Cd_3As_2$*, Phys. Rev. B **88**,125427 (2013).

14. Z. K. Liu, J. Jiang, B. Zhou, Z. J. Wang, Y. Zhang, H. M. Weng, D. Prabhakaran, S.-K. Mo, H. Peng, P. Dudin, T. Kim, M. Hoesch, Z. Fang, X. Dai, Z. X. Shen, D. L. Feng, Z. Hussain, and Y. L. Chen, *A Stable Three-Dimensional Topological Dirac Semimetal $Cd_3As_2$*, Nat. Mater. **13**, 677 (2014)

15. S. Borisenko, Q. Gibson, D. Evtushinsky, V. Zabolotnyy, B. Büchner, and R. J. Cava, *Experimental Realization of a Three-Dimensional Dirac Semimetal*, Phys. Rev. Lett. **113**, 027603 (2014).

16. M. Neupane, S.-Y. Xu, R. Sankar, N. Alidoust, G. Bian, C. Liu, I. Belopolski, T.-R. Chang, H.-T. Jeng, H. Lin, A. Bansil, F. Chou, and M. Z. Hasan, *Observation of a Three-Dimensional Topological Dirac Semimetal Phase in High-Mobility $Cd_3As_2$*, Nat. Commun. **5**, 3786 (2014).

17. H. B. Nielsen and M. Ninomiya, *Absence of Neutrinos on a Lattice: (I). Proof by homotopy theory*, Nucl. Phys. B **185**, 20 (1981).

18. H. B. Nielsen and M. Ninomiya, *Absence of Neutrinos on a Lattice: (II). Intuitive Topological Proof*, Nucl. Phys. B **193**, 173 (1981).

19. J. Von Neumann and E. Wigner, *Concerning the Behaviour of Eigenvalues in Adiabatic Processes*, Phys. Z. **30**, 467 (1929).

20. C. Herring, *Accidental Degeneracy in the Energy Bands of Crystals*, Phys. Rev. **52**, 365 (1937).

21. S. Murakami, *Phase transition between the Quantum Spin Hall and Insulator Phases in 3D: Emergence of a Topological Gapless Phase*, New J. Phys. **9**, 356 (2007).

22. S. Murakami, *Gap Closing and Universal Phase Diagrams in Topological Insulators*, Physica E **43**, 738 (2011).

23. S. Adler, *Axial-Vector Vertex in Spinor Electrodynamics*, Phys. Rev. **177**, 2426 (1969).

24. J. S. Bell and R. Jackiw, *A PCAC Puzzle: $\pi^0 \to \gamma\gamma$ in $\sigma$-Model*, Nuovo Cimento **60A**, 47 (1969).

25. H. B. Nielsen, M. Ninomiya, *The Adler-Bell-Jackiw Anomaly and Weyl Fermions in a Crystal*, Phys. Lett. B **130**, 389 (1983).

26. P. Hosur and X.-L. Qi, *Recent Developments in Transport Phenomena in Weyl Semimetals*, Comptes Rendus Physique **14**, 857 (2013).

27. S. A. Parameswaran, T. Grover D. A. Abanin, D. A. Pesin, and A. Vishwanath, *Probing the Chiral Anomaly with Nonlocal Transport in Three-Dimensional Topological Semimetals*, Phys. Rev. X **4**, 031035 (2014).

28. A. A. Burkov, *Chiral Anomaly and Transport in Weyl Metals*, J. Phys.: Cond. Matt. **27**, 113201 (2015).

29. A. A. Burkov and L. Balents, *Weyl Semimetal in a Topological Insulator Multilayer*, Phys. Rev. Lett. **107**, 127205 (2011).

30. G. B. Halász and L. Balents, *Time-Reversal Invariant Realization of the Weyl Semimetal Phase*, Phys. Rev. B **85**, 035103 (2012).

31. A. A. Zyuzin, S. Wu, A. A. Burkov, *Weyl Semimetal with Broken Time Reversal and Inversion Symmetries*, Phys. Rev. B **85**, 165110 (2012).

32. J. Liu and D. Vanderbilt, *Weyl Semimetals from Noncentrosymmetric Topological Insulators*, Phys. Rev. B **90**, 155316 (2014).

33. D. Bulmash, C.-X. Liu, and X.-L. Qi, *Prediction of a Weyl Semimetal in $Hg_{1-x-y}Cd_xMn_yTe$*, Phys. Rev. B **89**, 081106 (2014).

34. M. Hirayama, R. Okugawa, S. Ishibashi, S. Murakami, and T. Miyake, *Weyl Node and Spin Texture in Trigonal Tellurium and Selenium*, Phys. Rev. Lett. **114**, 206401 (2015).

35. H. Weng, C. Fang, Z. Fang, B. A. Bernevig, and X. Dai, *Weyl Semimetal Phase in Noncentrosymmetric Transition-Metal Monophosphides*, Phys. Rev. X **5**, 011029 (2015).

36. S.-M. Huang, S.-Y. Xu, I. Belopolski, C.-C. Lee, G. Chang, B. Wang, N. Alidoust, G. Bian, M. Neupane, C. Zhang, S. Jia, A. Bansil, H. Lin, and M. Z. Hasan, *A Weyl Fermion Semimetal with Surface Fermi Arcs in the Transition Metal Monopnictide TaAs Class*, Nat. Commun. **6**, 7373 (2015).

37. H. Weng, X. Dai, and Z. Fang, *Exploration and Prediction of Topological Electronic Materials Based on First-Principles Calculations*, MRS Bulletin **39**, 849 (2014).

38. B. Q. Lv, N. Xu, H. M. Weng, J. Z. Ma, P. Richard, X. C. Huang, L. X. Zhao, G. F. Chen, C. Matt, F. Bisti, V. Strokov, J. Mesot, Z. Fang, X. Dai, T. Qian, M. Shi, and H. Ding, *Observation of Weyl Nodes in TaAs*, arXiv:1503.09188.

39. S.-Y. Xu, I. Belopolski, N. Alidoust, M. Neupane, G. Bian, C. Zhang, R. Sankar, G. Chang, Z. Yuan, C.-C. Lee, S.-M. Huang, H. Zheng, J. Ma, D. S. Sanchez, B. Wang, A. Bansil, F. Chou,

P. P. Shibayev, H. Lin, S. Jia, and M. Z. Hasan, *Discovery of a Weyl Fermion Semimetal and Topological Fermi Arcs*, Science DOI:10.1126/science.aaa9297.

40. S. Furuseth, K. Selte, A. Kjekshus, *On the Arsenides and Antimonides of Tantalum*, Acta. Chem. Scand. **19**, 95 (1965).
41. http://www.openmx-square.org
42. J. P. Perdew, K. Burke, and M. Ernzerhof, *Generalized Gradient Approximation Made Simple*, Phys. Rev. Lett. **77**, 3865 (1996).

Fig. 1. (Color Online) Weyl semimetal and TaAs single crystal. (A) A 3D Dirac-cone point (DCP) can be driven into Weyl nodes with opposite chirality by either time-reversal symmetry breaking (TRB) or inversion-symmetry breaking (ISB). (B) A Weyl semimetal has Fermi arcs on its surface connecting projections of two Weyl nodes with opposite chirality. A Weyl node behaves as a magnetic monopole (MMP) in momentum space and its chirality corresponds to the charge of the MMP. (C) Crystal structure of TaAs. The arrow indicates that the cleavage occurs between the As and Ta atomic layers, producing two kinds of (001) surfaces with either As or Ta terminated atomic layers. (D) Bulk and projected (001) surface BZs with high-symmetry points indicated. (E) The core level photoemission spectrum measured with a photon energy $hv$ = 70 eV shows the characteristic As 3$d$, Ta 5$p$ and Ta 4$f$ peaks. (F) and (G) are pictures taken by metallographic microscopy of the cleaved surface of one TaAs sample used in our ARPES measurements. (H) ARPES spectra along $\bar{\Gamma}$-$\bar{X}$ showing highly dispersive and well-defined quasiparticle peaks.

Fig. 2. (Color Online) Electronic structure of (001) surface state. (A)-(D) Curvature intensity plots of ARPES data along $\bar{\Gamma}$-$\bar{X}$ at $hv$ = 30, 36, 42 and 48 eV, respectively. Dashed lines represent the extracted band dispersions. (E) Extracted band dispersions from the experimental data in (A)-(D). (F) Curvature intensity plot of ARPES data at $E_F$ along $\bar{\Gamma}$-$\bar{X}$ as a function of momentum and photon energy, for which the $k_z$ range at the BZ center is estimated to be 6.1 – 8.1 π/$c$' with the inner potential of 15 eV, which $c$' is one half of the $c$-axis lattice constant of TaAs. (G) Calculated band structure along $\bar{\Gamma}$-$\bar{X}$ for a seven-unit-cell-thick (001) slab with As and Ta terminations on the top and bottom surfaces, respectively. The intensity of the red color scales with the wave function spectral weight projected to one unit cell at the top surface with As termination. (H) Same as (G) but the spectral weight is projected to the bottom surface with Ta termination. (I) and (J) Intensity plot of ARPES data and corresponding curvature intensity plot along the high-symmetry lines $\bar{M}$-$\bar{X}$-$\bar{\Gamma}$-$\bar{Y}$-$\bar{M}$-$\bar{\Gamma}$, respectively. For comparison, the calculated surface bands on the As terminated surface are plotted on top of the experimental data in

(J). (K) Calculated band structure along $\overline{M}$ - $\overline{X}$ - $\overline{\Gamma}$ - $\overline{Y}$ - $\overline{M}$ - $\overline{\Gamma}$. The red color indicates the surface bands on the As terminated surface.

Fig. 3. (Color Online) Fermi surface on TaAs (001) surface. (A) Photoemission intensity plot at $E_F$ in the (001) surface recorded at $hv$ = 54 (outside the green box) and 36 eV (inside the green box). Larger and smaller circles indicate the locations of W1 and W2 projected onto the (001) surface BZ, respectively. Black and red colors represent opposite chirality. (B) Theoretical Fermi surface in (001) surface BZ. (C) Schematic of a closed Fermi surface pocket and an open Fermi arc crossing an arbitrary $k$-loop in the two-dimensional case. (D) Blue and magenta circles indicate the locations where the surface bands cross the enclosed $k$-loop $\overline{\Gamma}$ - $\overline{X}$ - $\overline{M}$ - $\overline{\Gamma}$ and $\overline{\Gamma}$ - $\overline{Y}$ - $\overline{M}$ - $\overline{\Gamma}$ at $E_F$, respectively. (E) and (F) Photoemission intensity map and energy distribution curves along $\overline{X}$ - $\overline{M}$ [$C1$ indicated in (A)], respectively. (G) and (H) Same as (E) and (F), respectively, but recorded along $C2$. (I) and (J) Same as (E) and (F), respectively, but recorded along $\overline{Y}$ - $\overline{M}$ [$C3$ indicated in (A)]. The calculated surface state bands along $\overline{X}$ - $\overline{M}$ and $\overline{Y}$ - $\overline{M}$ are plotted on top of the experimental data in (E) and (I), respectively. The dashed lines in (H) and (J) track the band dispersions.

Fig. 4. (Color Online) Surface Fermi arcs on TaAs (001) surface. (A) and (B) Fine $k$-point sampling calculations of surface states at $E_F$ around W2 and W1 along $\overline{\Gamma}$-$\overline{Y}$, respectively. W1 are indicated by dashed circles since the chemical potential is slightly away from the nodes. $a$1-$a$5 represent the arc-like lines connecting to W1. (C) Schematic of the Fermi arcs connecting pattern. Solid and hollow circles represent the projected Weyl nodes with opposite chirality and numbers inside indicate the total chiral charge. (D) and (E) Photoemission intensity plot at $E_F$ around W1 and extracted Fermi arcs, respectively. (F) Same as (B) but nearby $\overline{\Gamma}$ - $\overline{X}$. $b$1-$b$5 represent the arc-like lines connecting to W1. (G) and (H) Photoemission intensity plot and calculated surface state bands along $C4$ [indicated in (B)], respectively. (I) and (J) Same as (G) and (H), respectively, but recorded along $C5$ [indicated in (F)]. (K) Schematic of the locations of the cuts C6 and C7 near W2. (L) and (M) Curvature intensity plots of ARPES data along C6 ($k_y = 0.95\ \pi/a$) and C7 ($k_y = \pi/a$), respectively.

Fig. 5. (Color Online) (A)-(E) Curvature intensity plots of ARPES data along D1-D5, respectively. (F) ARPES intensity plot at $E_F$ around the Weyl-node projection W1. Red lines represent the momentum locations of cuts D1-D5 sliding through W1.

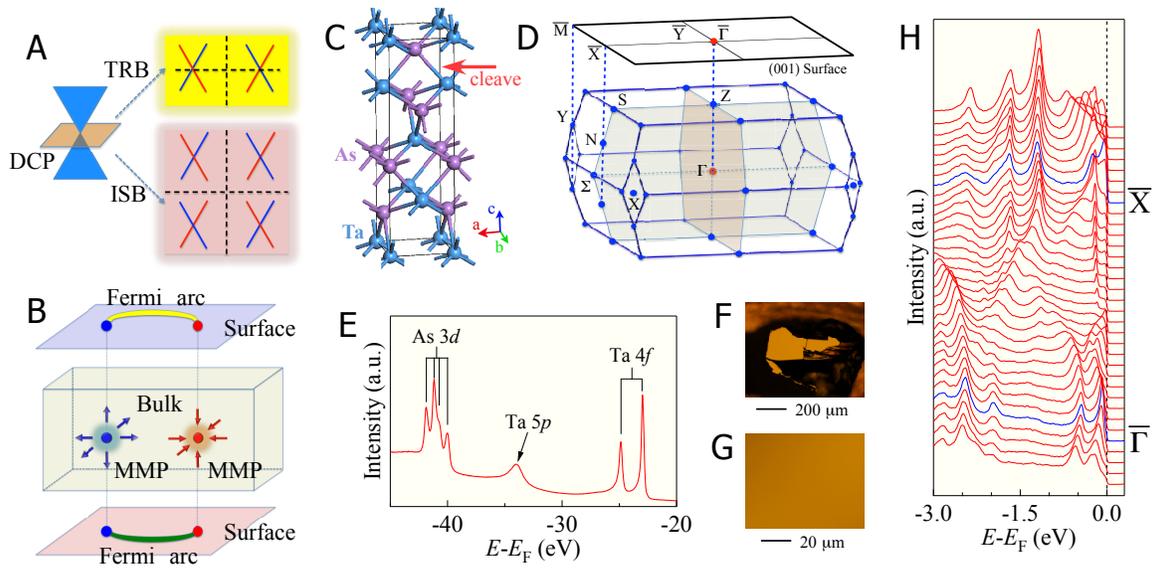

Fig. 1

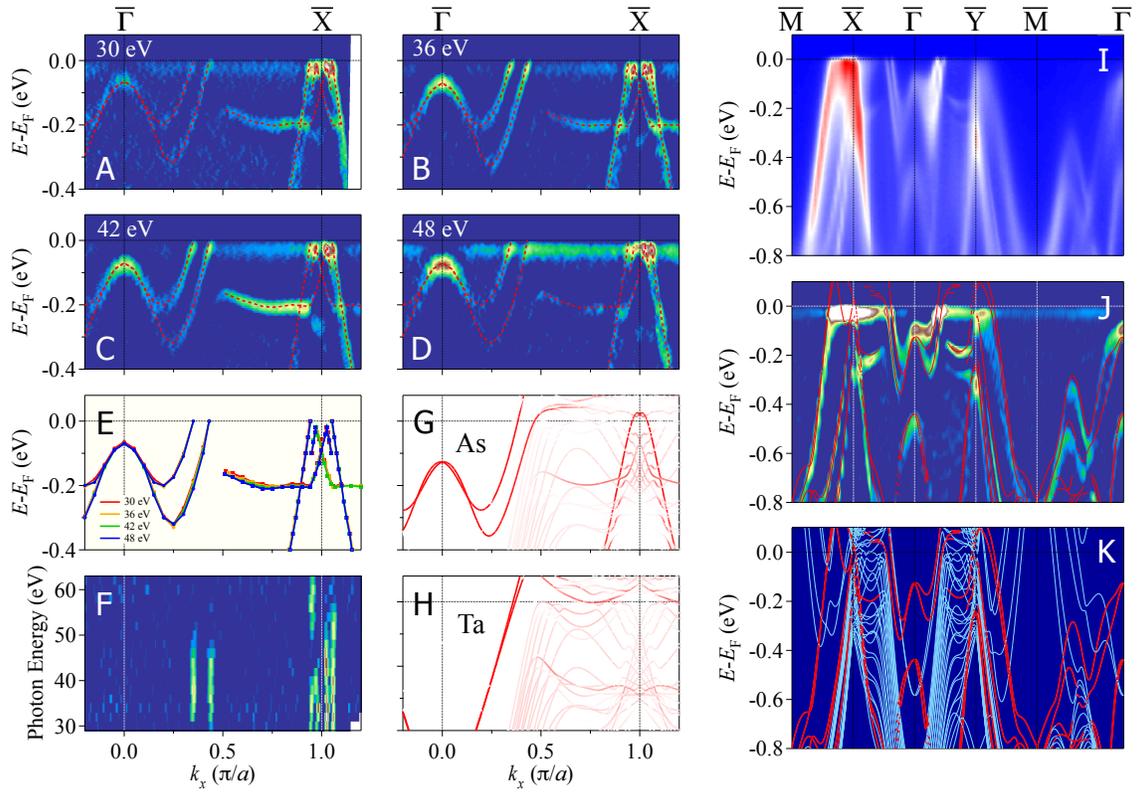

Fig. 2

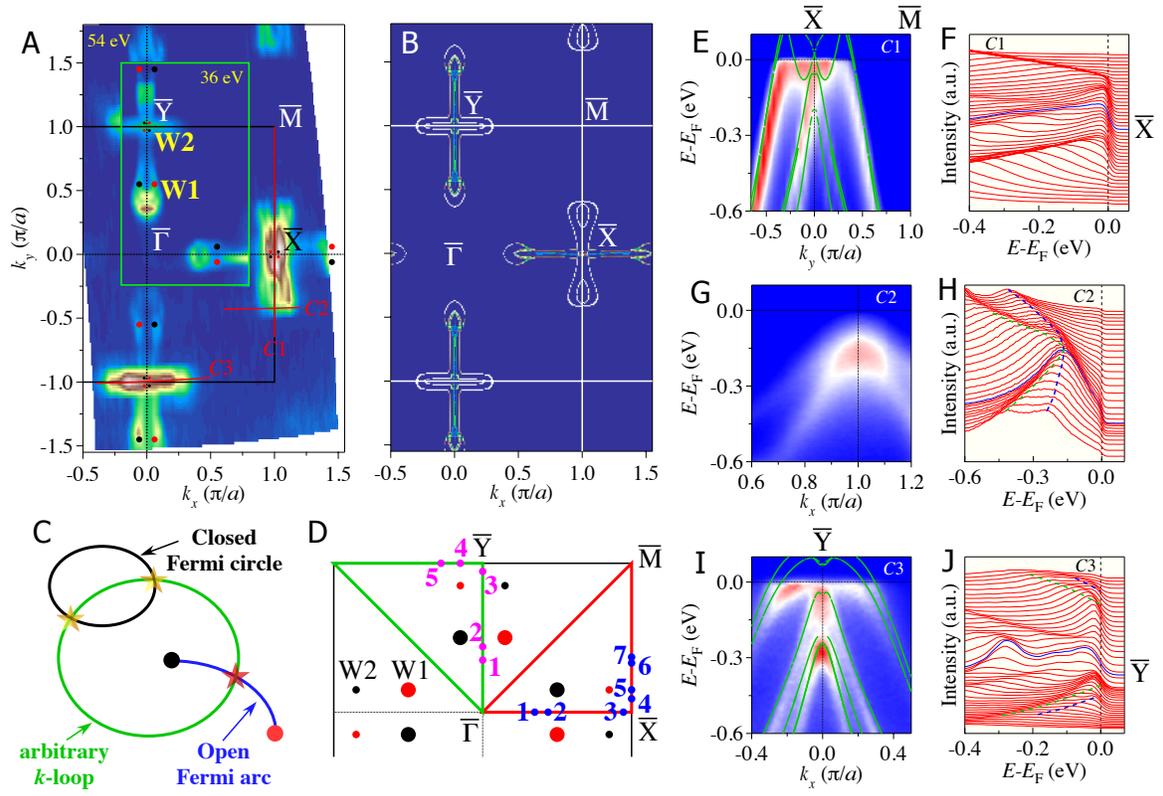

Fig. 3

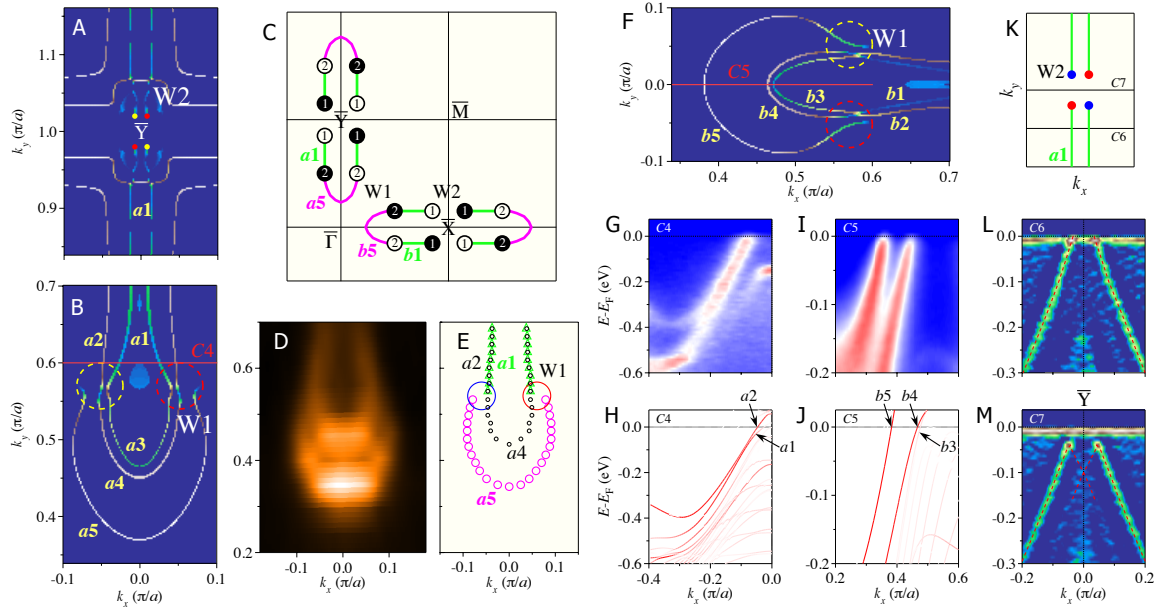

Fig. 4

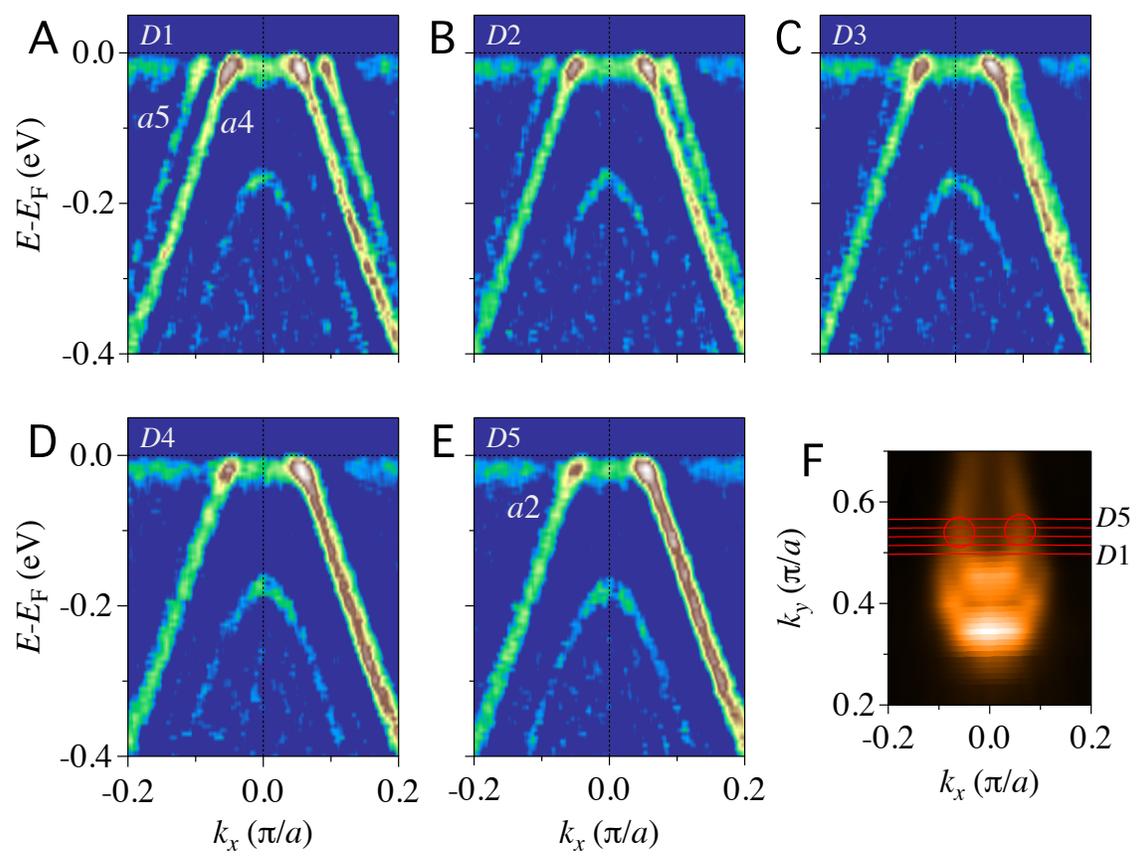

Fig. 5